 \documentclass{iopart}

\usepackage{amssymb}
\usepackage{bm}

\newcommand{\bra}[1]{\ensuremath{\bm{\langle}#1\bm{|}}}
\newcommand{\ket}[1]{\ensuremath{\bm{|}#1\bm{\rangle}}}

\begin{document}

\jl{3}

\title[``Unusual'' metals in two dimensions: \dots]{``Unusual'' metals in two dimensions:
one-particle model of the metal-insulator transition at $T=0$}

\author{Yu V Tarasov}

\address{Institute for Radiophysics and Electronics, National Academy of Sciences
of Ukraine, 12 Acad. Proskura St., Kharkov 61085, Ukraine}

\smallskip
\address{E-mail: \textit{yutarasov@ire.kharkov.ua}}

\begin{abstract}
The conductance of disordered nano-wires at $T=0$ is calculated in
one-particle approximation by reducing the original
multi-dimensional problem for an open bounded system to a set of
exactly one-dimensional non-Hermitian problems for mode
propagators. Regarding two-dimensional conductor as a limiting
case of three-dimensional disordered quantum waveguide, the
metallic ground state is shown to result from its multi-modeness.
On thinning the waveguide (in practice, e. g., by means of the
``pressing'' external electric field) the electron system
undergoes a continuous phase transition from metallic to
dielectric state. The result predicted conform qualitatively to
the observed anomalies of the resistance of different planar
electron and hole systems.
\end{abstract}

\pacs{71.30.+h, 72.15.Rn, 73.50.-h}


\nosections

In physics of disordered systems, there was a long-standing belief
initiated by scaling theory of localization~\cite{bib:AALR79} that
in two-dimensional (2D) electron and hole systems, just as in
one-dimensional ones (1D), there can be no metallic ground state
at any strength of disorder. However, numerous studies on
different materials performed in recent years (see extensive
bibliography in reference \cite{bib:AKS01}) have challenged this
point of view and showed it to be apparently incomplete and may
be, in general case, incorrect.

Inconsistency of the experimental facts with theoretical
predictions have led to a vast amount of publications where the
observed unusual phenomena were interpreted with the use of
different physical ideas. Among them were the formation of
conducting state in electron systems of very low density
\cite{bib:F84,bib:CCL98}, the existence of esoteric ideal metallic
state \cite{bib:DAMC97}, non-Fermi-liquid behaviour of the
conduction electrons~\cite{bib:CYA98}, different types of
superconductivity \cite{bib:BK98,bib:TN98}, the effects, which are
classical in nature, of traps formation at the interface of
slightly different materials \cite{bib:AM99}, and the temperature
dependent screening of electron-impurity scattering
\cite{bib:KS99,bib:SH00}.

In this paper, we focus our attention on interpreting the observed
phenomena in terms of the concept of ``quantum dephasing''. Within
the framework of this approach, the ``anomalous'' conducting state
of 2D systems is regarded as the result of phase randomization of
the originally coherent localized electrons due to their
interaction with some systems considered as the ``environment''.
The physical nature of dephasing environment in real systems still
remains controversial \cite{bib:AAG99}. As the most probable cause
of dephasing, some authors suggest quasi-elastic Coulomb
interaction of carriers since the ``anomalous'' behaviour of the
resistance is commonly observed in 2D systems with low electron
density ($r_s\gtrsim 10$). Note, however, that different theories
evaluate this kind of interaction quite differently, some as
promoting the localization~\cite{bib:AAL80,bib:TC89} whereas some
as inhibiting its origination~\cite{bib:F84,bib:CCL98,bib:BK94}.

Meanwhile, it has been recently shown \cite{bib:Tar00} that
scattering from static inhomogeneities can lead, much the same way
as different inelastic processes do, to the dephasing of quantum
states properly classified with regard for the confinement of the
dynamical system being considered. The arguments were based on the
use of the mode representation for one-particle propagators, which
seems to be most appropriate as applied to \emph{open systems} of
waveguide type. In regard to electrons (considered as quantum
waves), mode states represent the \emph{collective} excitations
which describe the electron system as strongly correlated. As a
matter of fact, the correlation, even without invoking Coulomb
interaction, is originally embedded in the Green function
formalism which explicitly takes into account the Pauli principle
\cite{bib:MA00}.

As was shown in \cite{bib:Tar00}, owing to dephasing of the
coherent mode states through impurity scattering of the electrons
between the \emph{extended} modes (for which the term ``open
channels'' is widely used) the formation of Anderson localized
states in two-dimensional conductors that are not extremely narrow
(i. e. possess more than one quantum mode) is made possible under
some particular, though realizable, conditions only. Meanwhile,
although from the results given in \cite{bib:Tar00} it follows
that the metallic-like ground state should not be considered to be
anomalous for 2D disordered systems, the analysis did not reveal
the clear mechanism for changing such systems over from the
conducting to insulating state and vice versa, which is observed
in numerous experiments (see the bibliography in
reference~\cite{bib:AKS01}).

In this paper, the method previously developed for exactly
two-dimensional open systems is extended to the case of conductors
of larger dimensionality in order to fit the formal statement of
the problem to real experimental conditions. The approach is
inspired by the fact that in practice 2D conducting systems are
created mostly by forming near-surface potential finite-width
wells. The wells are produced either due to application of
``pressing'' external electric field or due to the contact
potential. The shape of the well (in most cases it is nearly
triangular) is of no crucial importance for its principal
function, viz. to restrict the electron motion in the direction
normal to the heterogeneity surface. Therefore, below we examine
somewhat simplified model of the conductor which is chosen in the
form of a rectangular three-dimensional quantum waveguide with
``hard'' side boundaries. The length $L$ of the waveguide along
the $x$-axis as well as the width $W$, $y$-axis, and the height
$H$ in the direction of $z$-axis will be considered as arbitrary.
We assume the chemical potential (or the Fermi energy) to be
constant with change in the geometrical parameters. Such a model
is justified for an \emph{open} system, i. e. the conductor
attached to identical reservoirs in the equilibrium.

According to the linear response theory \cite{bib:K57}, the
conductance can be expressed in terms of the retarded one-particle
Green function of the electron system. This function, within the
model of isotropic Fermi liquid and with the units such that
$\hbar\hm=2m\hm=1$ ($m$~is the electron effective mass), satisfies
the equation
\begin{equation}\label{master}
  \left[
    \Delta+k_F^2+{\rm i}0-V(\bi{r})
  \right]
  G(\bi{r},\bi{r}')=\delta(\bi{r}-\bi{r}') \ .
\end{equation}
Here, $\Delta$ is the three-dimensional Laplacian, $k_F$ is the
electron Fermi wavenumber, $V(\bi{r})$ is the static random
potential specified by zero mean value, $\langle
V(\bi{r})\rangle=0$, and the binary correlation function
\begin{equation}\label{corr3d_simpl}
  \left<V(\bi{r})V(\bi{r}')\right>=\mathcal{QW}(x-x')
  \delta(\bi{r}_{\perp}-\bi{r}'_{\perp}) \ ,
\end{equation}
where $\bi{r}_{\perp}=(y,z)$. The function $\mathcal{W}(x)$ is
normalized to unity and decays at the scale of~$r_c$, the
correlation radius.

Equation (\ref{master}) must be supplemented with proper boundary
conditions. The ``impenetrable'' for quantum waves side boundaries
of the system can be characterized by a real impedance. On the
contrary, the open ends of the waveguide ($x=\pm L/2$) correspond
to a complex impedance owing to which the differential operation
in (\ref{master}) becomes non-Hermitian.

In paper \cite{bib:Tar00}, the method was proposed for solving
such a non-Hermitian problem in the case of two-dimensional
conducting systems. The analogous procedure is applicable to
waveguide-type systems of arbitrary dimensionality as well. The
essence of the method is the transition from one transport problem
posed in the dimension higher than unity to an infinite set of
exactly one-dimensional problems for mode Fourier-components of
the propagator $G(\bi{r},\bi{r}')$. For the waveguide-type system
under study, the transition to mode representation can be carried
out using the whole set of eigenfunctions
$\ket{\bi{r}_{\perp};\bm{\mu}}$ of the ``transverse'' Laplace
operator, which is made up of ordinary trigonometric functions
($\bm{\mu}=(n,m)$ is the vectorial mode index specified by
integer-valued components). With such an eigen set, equation
(\ref{master}) is transformed into a set of equations for mode
components $G_{\bm{\mu}\bm{\mu}'}(x,x')$ of the function
$G(\bi{r},\bi{r}')$,
\begin{equation}\label{mode_eq}
\fl  \left[\frac{\partial^2}{\partial x^2}+
  \kappa^2_{\bm{\mu}}+{\rm i}0-
  V_{\bm{\mu}}(x)\right]G_{\bm{\mu}\bm{\mu}'}(x,x')
  -\sum_{\bm{\nu}\neq\bm{\mu}}U_{\bm{\mu}\bm{\nu}}(x)
  G_{\bm{\nu}\bm{\mu}'}(x,x')=
  \delta_{\bm{\mu}\bm{\mu}'}\delta(x-x') \ .
\end{equation}
The parameter
\begin{equation}\label{kappa_nm}
  \kappa^2_{\bm{\mu}}=k_F^2-\left({\pi n}/{W}\right)^2-
  \left({\pi m}/{H}\right)^2
\end{equation}
in (\ref{mode_eq}) has the meaning of a \emph{longitudinal energy}
of the mode $\bm{\mu}$, the potential matrix
$\|U_{\bm{\mu}\bm{\mu}'}\|$ is composed of the functions
\begin{equation}\label{Umatr}
  U_{\bm{\mu}\bm{\mu}'}(x)=\int_S\textrm{d}\bi{r}_{\perp}
  \ket{\bi{r}_{\perp};\bm{\mu}}
  V(\bi{r})\bra{\bi{r}_{\perp};\bm{\mu}'} \ .
\end{equation}
The diagonal components of this matrix, $V_{\bm{\mu}}(x)\equiv
U_{\bm{\mu}\bm{\mu}}(x)$, are responsible for the
\emph{intra-mode} whereas off-diagonal components --- for the
\emph{inter-mode} scattering of quantum waves.

The initial problem reformulated in terms of the
``one-coordinate'' differential equations (\ref{mode_eq}) cannot
actually be considered as strictly one-dimensional for the
entanglement of overall mode components of the Green function
matrix $\|G_{\bm{\nu}\bm{\mu}}\|$. In reference~\cite{bib:Tar00},
however, it was shown that all the off-diagonal elements of this
matrix can be expressed, by means of a linear operation, in terms
of the corresponding diagonal elements,
\begin{equation}\label{G_mn->G_nn}
  G_{\bm{\nu}\bm{\mu}}(x,x')=\int_L\mathrm{d}x_1\,
  \mathsf K_{\bm{\nu}\bm{\mu}}(x,x_1)G_{\bm{\mu}\bm{\mu}}(x_1,x') \
  ,\hspace{1cm} \bm{\nu}\neq\bm {\mu} \ .
\end{equation}
The kernel of the operator (\ref{G_mn->G_nn}) can be determined
exactly given the solution $G^{(V)}_{\bm{\mu}}(x,x')$ of the
truncated equation~(\ref{mode_eq}),
\begin{equation}\label{trial}
  \left[\frac{\partial^2}{\partial x^2}+
  \kappa^2_{\bm{\mu}}+{\rm i}0-
  V_{\bm{\mu}}(x)\right]G^{(V)}_{\bm{\mu}}(x,x')
  =\delta(x-x') \ ,
\end{equation}
i.~e. the equation with all inter-mode potentials
$U_{\bm{\mu}\bm{\nu}}(x)$  omitted. For (\ref{trial}), the
requirement of the waveguide openness at the ends $x=\pm L/2$ can
be formulated in terms of Sommerfeld's radiation conditions
\cite{bib:BF,bib:V67},
\begin{equation}\label{rad_cond}
  \left(\frac{\partial}{\partial x}\mp {\rm
  i}\kappa_{\bm{\mu}}\right)
  G^{(V)}_{\bm{\mu}}(x,x')\Bigg|_{x=\pm L/2}=0 \ ,
\end{equation}
which correspond to adiabatic (not resulting in scattering)
attachment of the conductor at hand to the ideally conducting
leads.

The auxiliary ``trial'' Green function $G^{(V)}_{\bm{\mu}}(x,x')$,
in view of statistical formulation of the problem, can be thought
of as known precisely if one manages to find all its statistical
moments $\langle\big[G^{(V)}_{\bm{\mu}}(x,x')\big]^{p}\rangle$,
$p\in\aleph$. Using the technique described in \cite{bib:Tar00},
these moments can be obtained provided the scattering from the
potential $V(\bi{r})$ is regarded as weak. The weakness criteria
can be formulated in terms of the characteristic lengths,
\begin{equation}\label{weakness}
  k_F,\,r_c\ll\ell\ ,
\end{equation}
where $\ell$ stands for the quasi-classical mean-free path of
conducting electrons. Its value, in the particular case of
completely $\delta$-correlated random potential
($\mathcal{W}(x)=\delta(x)$ in (\ref{corr3d_simpl})), is equal to
$4\pi/\mathcal{Q}$.

For the extended modes ($\kappa^2_{\bm{\mu}}>0$), calculation of
the required moments yields
\begin{eqnarray}\label{Gv_moments}
\fl \Big<\left[G^{(V)}_{\bm{\mu}}(x,x')\right]^{p}\Big>
  =\left(\frac{-\mathrm{i}}{2\kappa_{\bm{\mu}}}\right)^{p}
  \exp\Bigg[  \mathrm{i}p & \kappa_{\bm{\mu}}|x-x'|  \nonumber\\
&  -\frac{p}{2}
  \left(\frac{p}{L_f^{(V)}(\bm{\mu})}+
  \frac{1}{L_b^{(V)}(\bm{\mu})}\right)|x-x'|\Bigg] \ . 
\end{eqnarray}
Here, $L_{f,b}^{(V)}(\bm{\mu})$ are the forward ($f$) and the
backward ($b$) mode scattering lengths associated solely with the
\emph{intra-mode} potential $V_{\bm{\mu}}(x)$,
\begin{equation}\label{ext_length}
  L_f^{(V)}(\bm{\mu})=\frac{4S}{9\mathcal{Q}}
  \left(2\kappa_{\bm{\mu}}\right)^2 \ ,\hspace{1cm}
  L_b^{(V)}(\bm{\mu})=\frac{4S}{9\mathcal{Q}}
  \frac{\left(2\kappa_{\bm{\mu}}\right)^2}%
  {\widetilde{\mathcal{W}}(\kappa_{\bm{\mu}})} \ ,
\end{equation}
$S$ is the conductor cross-section area,
$\widetilde{\mathcal{W}}(\kappa_{\bm{\mu}})$ is the Fourier
transform of $\mathcal{W}(x)$. As far as the evanescent modes are
concerned ($\kappa^2_{\bm{\mu}}<0$), at weak scattering the
potential $V_{\bm{\mu}}(x)$ in equation (\ref{trial}) can be
omitted, thus allowing one to take advantage of the unperturbed
solution,
\begin{equation}\label{evan}
  G^{(V)}_{\bm{\mu}}(x,x')=-\frac{1}{2|\kappa_{\bm{\mu}|}}
  \exp\big(-|\kappa_{\bm{\mu}}||x-x'|\big) \ .
\end{equation}

It is worth noting that if the particular functions
$G^{(V)}_{\bm{\mu}}(x,x')$ were used instead of the precise mode
propagators, that is if for some reasons the inter-mode scattering
is neglected \footnote[1]{In some particular cases, e. g. if the
conductor inhomogeneity is in the form of random strata along the
$x$-axis, all the intermode potentials vanish exactly.}, the
result widely known from the theory of quasi-1D conductors would
be obtained. Specifically, the conductance would
\emph{exponentially} decrease as the conductor length grows when
the latter exceeds the value of the order of $N_c\ell$ (where
$N_c$ is the number of open channels). It is exactly this length
scale that was associated previously with the localization length
in disordered 2D systems \cite{bib:D84,bib:MPK88}.

However, in general case, there are no grounds for neglecting the
inter-mode potentials in equation (\ref{mode_eq}). Taking them
into account leads to the multi-channel Lippmann-Schwinger
equation \cite{bib:T75} for the kernel of the operator
(\ref{G_mn->G_nn}), which can be explicitly solved if the Green
function of trial equation (\ref{trial}) is chosen as a starting
approximation for the exact mode propagator. As a result, the
closed one-dimensional equation for the diagonal component
$G_{\bm{\mu}\bm{\mu}}$ is deduced,
\begin{equation}\label{GDIAG-FIN}
  \left[\frac{\partial^2}{\partial x^2}+
  \kappa^2_{\bm{\mu}}+{\rm i}0-V_{\bm{\mu}}(x)-\hat{\mathcal T}_{\bm{\mu}}\right]
  G_{\bm{\mu}\bm{\mu}}(x,x')=\delta(x-x') \ .
\end{equation}
Here, $\hat{\mathcal T}_{\bm{\mu}}$ is the non-local operator
potential which, with the proviso of (\ref{weakness}), equals to
$\hat{\mathcal T}_{\bm{\mu}}\approx\bm{P}_{\bm{\mu}}\hat{\mathcal
U}\hat{G}^{(V)}\hat{\mathcal U}\bm{P}_{\bm{\mu}}$. The operators
$\hat{\mathcal U}$ and $\hat{G}^{(V)}$ are defined in the mixed
coordinate-mode subspace $(x,\bm{\nu})$ with the excluded mode
$\bm{\mu}$ and specified by the matrix elements
\begin{eqnarray}
  &\ket{x,\bm{\nu}}\hat{\mathcal U}\bra{x',\bm{\nu}'} =\,
  U_{\bm{\nu}\bm{\nu}'}(x)\delta(x-x') \label{Umat}\ ,\\
\bs  &\ket{x,\bm{\nu}}\hat{G}^{(V)}\bra{x',\bm{\nu}'} =\,
  G^{(V)}_{\bm{\nu}}(x,x')\delta_{\bm{\nu}\bm{\nu}'} \ ,
\label{Gmat}
\end{eqnarray}
$\bm{P}_{\bm{\mu}}$ is the projection operator from the above
pointed subspace into the mode $\bm{\mu}$. Equation
(\ref{GDIAG-FIN}) along with (\ref{G_mn->G_nn}) enables one to
obtain the function $G(\bi{r},\bi{r}')$ with the required accuracy
and in so doing to determine the system conductance.

If there is more than one open channel in the conductor (at least,
two), the operator potential $\hat{\mathcal T}_{\bm{\mu}}$ results
in appearing the imaginary term $i/\tau_{\bm{\mu}}^{(\varphi)}$ in
the spectrum of (\ref{GDIAG-FIN}),
\begin{equation}\label{dephase}
  \frac{1}{\tau_{\bm{\mu}}^{(\varphi)}} =
  \frac{\mathcal{Q}}{4S}{\sum_{\bm{\nu}\neq\bm{\mu}}}' \frac{1}{\kappa_{\bm{\nu}}}
  \left[\widetilde{\mathcal{W}}(\kappa_{\bm{\mu}}-\kappa_{\bm{\nu}})+
  \widetilde{\mathcal{W}}(\kappa_{\bm{\mu}}+\kappa_{\bm{\nu}})\right]
  \approx \frac{k_F}{2\ell},
\end{equation}
which indicates the phase uncertainty arisen owing to the
scattering between extended modes other than the particular mode
$\bm{\mu}$. In (\ref{dephase}), the prime indicates the summation
over \emph{open channels} only. This means that in systems with
more than one conducting channel the inter-mode scattering from
quenched disorder produces the same dephasing effect on the
many-particle carrier states as the real inelastic scattering
processes do. Still, there is some distinction consisting in that
scattering from static inhomogeneities does not affect actual
one-electron energy (at $T=0$ the latter remains the Fermi energy)
but it does influence the longitudinal energy of a many-particle
state, i. e. the quantum (or waveguide) mode.

As a result, on increasing the length $L$ of the multi-channel
conductor the conductance does not fall exponentially, which is
characteristic of Anderson-localized states, even if $L>N_c\ell$.
In particular, at $N_c\gg\nolinebreak 1$, the dimensionless (in
units of $e^2/\pi\hbar$) average conductance $\big<g(L)\big>$ is
given as
\numparts
\begin{eqnarray}
 \hspace{-1cm} \big< g(L)\big>\approx N_c\ ,\hspace{2.8cm}&&  L<\ell,
  \label{g_fin_ball}\\
  \hspace{-1cm} \big< g(L)\big>\approx (4/3)N_c\ell/L\ ,\hspace{1.4cm}&&  \ell\ll L \ .
  \label{g_fin_loc}
\end{eqnarray}
\endnumparts

If there is only one open channel in the conductor (with, say,
mode index $\bm{\mu}_1$) then there are no terms in the sum of
(\ref{dephase}) and, therefore, the electron state in this channel
should be thought of as coherent. In this case, the conductance is
obtained from the \emph{one-dimensional localization} theory. It
varies with $L$ according to the law
\begin{equation}\label{cases}
  \langle g(L) \rangle \approx
\cases{
  1-4L/\xi_1\ , &$\ L/\xi_1\ll 1 $\\
  A\left(\xi_1/L\right)^{3/2}
  \exp\left(-L/\xi_1\right)\ ,&$\ L/\xi_1\gg 1 $
}\ ,
\end{equation}
where $A=\pi^{5/2}/16$, $\xi_1=4L_b^{(V)}(\bm{\mu}_1)$. The result
(\ref{cases}) clearly indicates the localized character of the
electron states whose spatial extent, $\xi_1$, is determined by
the extended mode \emph{backscattering}.

Let us discuss the obtained results from the viewpoint of their
connection to the metal-insulator transition (MIT) observed in 2D
systems. First of all, it is worth noting that there are no signs
of MIT as long as the conductor is multi-mode. The result
(\ref{g_fin_ball}) is consistent with the conception of
\emph{ballistic} electron transport in short wires. In this case,
the conductance exhibits evident staircase-like dependence on the
electron energy as well as on the transverse dimension of the
conductor (for a massive three-dimensional wire the equality holds
$N_c=\nolinebreak\left[k_F^2S/4\pi\right]$, where $[\ldots]$
symbolizes the integer part of the number enclosed).

The asymptotic value (\ref{g_fin_loc}) is consistent with the
model of \emph{diffusive} classical motion of electrons and
coincides exactly with the conductance well-known from the kinetic
theory \footnote[2]{In this paper, we do not touch upon \emph{weak
localization} effects originating from the coherent scatterinng
due to the intra-mode potential $V_{\bm{\mu}}(x)$ in
(\ref{GDIAG-FIN}).}. The lack of the exponential dependence on $L$
within the length region $L\gg N_c\ell$ indicates the nonexistence
of the ``localized'' transport of current carriers in multi-mode
conductors. Although the original disorder (potential $V(\bi{r})$)
is of quenched character, the $T$-matrix $\hat{\mathcal
T}_{\bm{\mu}}$ in (\ref{GDIAG-FIN}) describes the recurrent
scattering of the $\bm{\mu}$-mode electrons through intermediate
states with mode energies different from $\kappa^2_{\bm{\mu}}$.
This scattering, once proceeding through the \emph{extended}
virtual modes, dephases ``parent'' mode state $\bm{\mu}$ quite
similarly to the actual inelastic scattering.

Now consider again the equation (\ref{GDIAG-FIN}) and the
expression (\ref{kappa_nm}) for the mode energy. Clearly, the
number of extended modes can be varied by changing the quantum
waveguide thickness only, keeping its width $W$ constant. By
decreasing $H$, one can gradually reduce the number of open
channels to unity, thus passing from ``metallic'' conductance,
either in its ballistic form (\ref{g_fin_ball}) or diffusive form
(\ref{g_fin_loc}), to ``dielectric'' result (\ref{cases}). In
practice, for the most part, by monitoring the shape of
near-surface potential wells the carrier density is normally
varied \cite{bib:BK94}, as a function of which the resistance of
2D systems is commonly measured.

On a further increase of the ``pressing'' voltage (accompanied by
the corresponding decrease of 2D carrier density) the quantum
waveguide turns into the below-cutoff one in that there remain
\emph{evanescent} modes only ($\kappa^2_{\bm{\mu}}<0$). These
states, as can be readily seen from (\ref{evan}), are
exponentially localized on a scale of the mode wavelength even in
the absence of any disorder. Naturally, as the electron system
enters into such a strongly localized phase, the resistance must
show a rapid increase (MIT).

To conclude, note that in most of the experimental works the
conclusions about transport properties of 2D electron and hole
systems were made on the basis of examining the temperature
dependence of their resistance. The detailed analysis of this
issue is beyond the scope of the present article. Nevertheless,
taking into account the above pointed features of
(quasi)-2D-systems ground state, a set of qualitative predictions
regarding the temperature dependence of their resistance can also
be made.

First note that the transition from the metallic-type conductance
(\ref{g_fin_ball}), (\ref{g_fin_loc}) to its small value in the
localized phase proceeds inevitably through the \emph{one-mode}
state of the electron system. It is evident from
(\ref{g_fin_ball}) and (\ref{cases}) that in such a transitional
regime the conductance should assume a value close to one
conductance quantum, irrespective of the material of the
conductor. It is exactly such a value that is observed in the
vicinity of the so-called ``separatrix'' in $T$-dependence of 2D
system resistivity \cite{bib:AKS01}. Also, the weak temperature
dependence of the separatrix can be explained if one takes into
account that mode wavelengths of marginal channels, viz. those
which are in the state of opening or closing, are large as
compared to the thermal phonon characteristic wavelength. If this
is the case, the interaction between those channels and the phonon
subsystem of the conductor should be ineffective.

In the insulating phase, where all of the modes are strongly
localized, it is natural to anticipate the temperature dependence
predicted by the percolation theory~\cite{bib:SE79}. Much more
intricate is the puzzle of non-monotonic dependence on $T$ of the
resistivity in the metallic phase in proximity to the separatrix.
Recall, however, that on approaching the separatrix from the
metallic side the electron system is certain to run through the
one-mode stage, i. e. it becomes virtually one-dimensional even
though the width of the quantum waveguide remains macroscopic. As
for 1D electron systems, non-monotonic temperature dependence of
their conductivity had been already predicted before, in reference
\cite{bib:AR78} for the case $T\tau\gg 1$ and in \cite{bib:T92}
for $T\tau\ll 1$.

Hence, the result obtained in this paper with the method of
reference~\cite{bib:Tar00}, which was extended to analyze real
two-dimensional electron systems, seems to be consistent with the
``unexpected'' experimental data extracted at low but non-zero
temperatures. This makes it reasonable to interpret the observed
peculiarities of the planar system resistance as revealing the
true \emph{continuous phase transition} \cite{bib:BK94,bib:SGCS97}
of such systems from conducting to insulating state.


\end{document}